\documentclass[12pt]{iopart}

\usepackage{amsfonts}
\usepackage{amssymb}
\usepackage[dvips]{graphicx}

\usepackage{bbm}
 
\begin{document}
 
 
\title{BCS model of Cooper Pair Box}
 
\author{Robert Alicki}

\address{  
Institute of Theoretical Physics and Astrophysics, University
of Gda\'nsk,  Wita Stwosza 57, PL 80-952 Gda\'nsk, Poland}
\ead{fizra@univ.gda.pl}
 
\begin{abstract}
The standard phenomenological Hamiltonian of a small superconducting Josephson junction in the charge regime (Cooper Pair Box)
produces a model of the effective \emph{charge qubit} with possible applications to  quantum information processing.
In this note a new model based on the BCS Hamiltonian with \emph{individual tunneling} yields an effective multi-level picture with a highly degenerated level
placed between the ground state and the excited state. Unlike in the standard approach, the excited Cooper pairs play here an important role. For such a system coupled to a zero temperature  bath the additional levels act as a probability sink. In contrast to the standard large-spin model the coupling to phonons can be an effective source of dissipation. This model provides also alternative explanations of various effects observed in  experiments and sheds new light on the issue of Josephson junctions as \emph{macroscopic quantum systems}.
\end{abstract}



\date{\today}
\maketitle
 
In the last decade remarkable experiments were performed involving measurements and manipulations of states for  a single or several nanoscopic Josephson junctions (JJ) which were consistently interpreted in terms of  two-level quantum systems \cite{W, Cla}. The type of the JJ considered here, the Cooper Pair Box (CPB), is a circuit consisting of a small superconducting island coupled via Josephson junction to a large superconducting reservoir \cite{Nak, Leh, Gui, Bla} . One of the standard forms of the CPB Hamiltonian is the Josephson Hamiltonian
\begin{equation}
\hat{H} = 4E_C( \hat{\mathrm{J}}_z - n_g)^2  - \frac{E_J}{2j}\hat{\mathrm{J}}_x .
\label{Jham}
\end{equation}
Here, $\hat{\mathrm{J}}_k, k=x,y,z$ are $j$-spin operators, where $j/2$ is a an average total number of Cooper pairs on the island at equilibrium. The first term describes the Coulomb repulsion and the parameter $n_g$ is an external control proportional to the gate voltage. The second term accounts for the tunneling of Cooper pairs between the island and the large electrode with the magnitude given by  the Josephson energy  $E_J < E_C $. At temperatures $k_B T << E_C$, and putting $0< n_g < 1$, it is enough
to restrict the analysis to the two lowest Coulomb energy states, say $|m=0\rangle , |m=1\rangle$, (for an integer spin) what leads to the picture of \emph{charge qubit}.
\par
Although there exist derivation of the Hamiltonian (\ref{Jham}) mainly based on the formal quantization of the phenomenological circuit equations or the two-mode approximation for the Bose-Einstein condensate(BEC) of Cooper pairs, the more rigorous approach is still needed. The main problem with the single degree of freedom model is its collective coupling to environment which should lead to a semiclassical behavior with normal fluctuations $\sim\sqrt{j}$ of the Cooper pairs number \cite{Ben}. This effect is observed in atomic BEC systems \cite{Est} which are described by the same Hamiltonian (\ref{Jham}) but clearly not in CPBs where charge fluctuations are strongly suppressed. In this note a solution of this problem is proposed, which is based on  the \emph{individual tunneling model} combined with the \emph{individual coupling} to a phonon's bath at zero temperature. The detailed derivations of the model Hamiltonians and the analysis of Markovian master equations, both for CPB and the \emph{current biased junction} (phase qubit) will be presented in the forthcoming publication \cite{AM}.
\par
\emph{Reduced BCS Hamiltonian}
In the following a more fundamental approach is used, which is based of the simplified BCS model \cite{BCS,Thou} which captures all essential features of superconductivity and  should be also valid for small JJ's including CPB.\\
In the absence of tunneling the island can be described by a system of Cooper pairs  treated as \emph{hard-core bosons} with the BCS mean-field interaction.
Equivalently, one can consider a system of $K$ spins-$1/2$ with spin operators $\hat{s}^{\alpha}_k$, $\alpha = x,y,z$ satisfying
\begin{equation}
[\hat{s}^x_k ,\hat{s}^y_l] = i\delta_{kl}\hat{s}^z_k ~~~ \mathrm{and\ cyclic\ permutations}.
\label{spin}
\end{equation}
Defining the collective spin  operators
\begin{equation}
\hat{\mathbf{J}}= (\hat{J}_x ,\hat{J}_y ,\hat{J}_z ),~~~\hat{J}_{\alpha}= \sum_{k=1}^K\hat{s}^{\alpha}_k\ ,\ \alpha = x,y,z,
\label{spincoll}
\end{equation}
one can write the mean-field BCS Hamiltonian
\begin{equation}
\hat{H}_{\mathrm{red}}= - \frac{g}{K}\sum_{k,l=1}^K \hat{s}^+_k\hat{s}^-_l = - \frac{g}{K} \bigl(\hat{\mathbf{J}}^2 - \hat{J}_z^2 + \hat{J}_z\bigr).
\label{bcs1}
\end{equation}
Here, $\hat{s}^+_k =\hat{s}^x_k +i\hat{s}^y_k, \hat{s}^-_l=\hat{s}^x_l -i\hat{s}^y_l$ create and annihilate a Cooper pair, $K$ is an even number of electron levels in the cut-off region around the Fermi energy and $g$ is an effective pairing potential.
Using the decomposition of the Hilbert space for $K$ spins-$1/2$ into subspaces corresponding to the irreducible representations of $SU(2)$ of the dimension $2j+1$ and multiplicity $r_j$
\begin{equation}
\mathbb{C}^{2^K} = \bigoplus_{j=0}^{K/2} \mathbb{C}^{2j+1}\otimes\mathbb{C}^{r_j}
\label{hilbert}
\end{equation}
one can use as eigenvectors of $\hat{H}_{\mathrm{red}}$ the orthonormal basis $|j,m;r\rangle$
\begin{equation}
\mathbf{J}^2|j,m;r\rangle= j(j+1)|j,m;r\rangle,~~~\hat{J}_z|j,m;r\rangle = m|j,m;r\rangle 
\label{basis}
\end{equation}
where $m = -j, -j+1,\ldots,j,\ r= 1,2,...,r_j.$
One should notice that the collective spin operators (\ref{spincoll}) do not correspond to a single spin, like in the case of the Josephson Hamiltonian (\ref{Jham}) but are direct sums of
$j$-spin operators with $j= K/2 , K/2-1,...,0$, and  with many copies for a given $j < K/2$.
The states with $j= K/2 - p$ describe excitations composed of $p$ \emph{excited Cooper pairs}.
In contrast to unpaired electrons this type of excitations are not common in the literature.
The main message of this paper is to argue that they play a crucial role in the description of
CPB.
\par
As at zero temperature the energy levels up to Fermi level are filled the number of Copper pairs $(K/2 + m)$  determined by the quantum number $m$ is close to $K/2$ and hence $m \simeq 0$. For the finite temperature case one obtains a superconducting phase transition at the critical temperature $k_B T_c = g/2$ \cite{Thou}. For a fixed $m$ the energy difference between the unique ground state $|K/2, m\rangle$ and the $(K-1)$-degenerated first excited level $|(K/2)-1, m ;r\rangle$ is equal to $g$ what implies that $g$ has a meaning of the zero temperature superconducting gap ($g\simeq \Delta$).

\par
\emph{Cooper Pair Box - BCS Hamiltonian}
 
Adding to the Hamiltonian $\hat{H}_{\mathrm{red}}$ the term describing Coulomb repulsion one obtains the Hamiltonian of the isolated island
\begin{equation}
\hat{H} =  - \frac{g}{K} \bigl(\hat{\mathbf{J}}^2 - \hat{J}_z^2 + \hat{J}_z\bigr)+ 4E_C \bigl(\hat{J}_z - \bar{m}\bigr)^2
\label{hamcpb2}
\end{equation}
where the charging energy $E_C$ fulfills $k_B T << E_C $ and $\bar{m}$ ( $m_0 \leq\bar{m}\leq m_0 +1, |m_0| << K$) determines the number of Cooper pairs in the systems,  and can be controlled by an external voltage. For experimental realizations the parameters $g \simeq \Delta \simeq k_B T_c$ and  $E_C$ are of the same order of magnitude ($\sim 1$ \emph{kelvin}), while $K\simeq 10^4 - 10^6$.
\par
Due to the \emph{Coulomb blockade} described by the second term in (\ref{hamcpb2}) one can consider only the states with $m=m_0, m_0 +1 $ satisfying $\bar{m}\in [m_0 , m_0 +1]$. Taking into account that $m_0 << K$ and extracting the irrelevant common constant $2E_C+4E_C [(m_0-\bar{m})^2+ (m_0-\bar{m})]- gK/4$ we have an effective Hilbert space spanned by the following vectors with the corresponding energies
denoted by the simplified symbols
\begin{equation}
\begin{array}{ll}
|0\rangle \equiv |K/2,m_0\rangle         &\ E_0 = -2E_C (1- 2n_g),\\
|1\rangle\equiv |K/2,m_0 +1\rangle        &\ E_1 = 2E_C [1-2n_g - 4\tilde{g}],\\
|s;0\rangle \equiv |K/2-1,m_0;s\rangle     &\ W_0 = -2E_C [1-2n_g - 4\tilde{g}],\\
|r;1\rangle\equiv |K/2-1,m_0+1;r\rangle   &\ W_1 = 2E_C (1- 2n_g)
\end{array}
\label{levels}
\end{equation}
where $s,r = 1,2,\ldots,K-1$ and
\begin{equation}
n_g = \bar{m} - m_0 - \tilde{g}\ ,\  \tilde{g} = \frac{g}{8E_C}.
\label{levels1}
\end{equation}

\emph{Tunneling processes}
 
Tunneling process for Cooper pairs produces superpositions of  states on the island which differ by a single pair and can be described by the  reduced tunneling Hamiltonian acting on the states of the island
\begin{equation}
\hat{T}_{\mathrm{red}} =  \frac{1}{2}\sum_{k=1}^K (\beta_k\hat{s}^+_k + \bar{\beta}_k\hat{s}^-_k)
\label{tunhamred}
\end{equation}
with, generally complex, tunneling probability amplitudes $\beta_k$.
Introducing the total amplitude $\beta=  \sum_{k=1}^K \beta_k $
one can decompose the tunneling Hamiltonian into collective and individual parts
\begin{equation}
\hat{T}_{\mathrm{red}} = \frac{1}{K}\bigl(\mathrm{Re}(\beta)\hat{J}_{x} +\mathrm{Im}(\beta)\hat{J}_{y}\bigr) +\frac{1}{2}\sum_{k=1}^K \Bigl[\bigl(\beta_k -\frac{1}{K}\beta\bigr)\hat{s}^+_k + \bigl(\bar{\beta}_k -\frac{1}{K}\bar{\beta}\bigr)\hat{s}^-_k\Bigr].
\label{tunham1}
\end{equation}
The collective part of (\ref{tunham1})  preserves the subspaces of a given $j$. Therefore,   if the collective part dominates one could consider only the states with $j = K/2$ to obtain the standard picture of a large spin with the Hamiltonian (\ref{Jham}). Indeed, the obtained collective Hamiltonian
\begin{equation}
\hat{H}^c =  - \frac{g}{K} \bigl(\hat{\mathbf{J}}^2 - \hat{J}_z^2 + \hat{J}_z\bigr)+ 4E_C \bigl(\hat{J}_z - \bar{m}\bigr)^2 + \frac{1}{K}\bigl(\mathrm{Re}(\beta)\hat{J}_{x} +\mathrm{Im}(\beta)\hat{J}_{y}\bigr)
\label{hamcpb1}
\end{equation}
restricted to the $j = K/2, |m| << K$ subspace and in the limit of large $K$ is unitarily equivalent (up to the irrelevant constant) to the large spin Hamiltonian (\ref{Jham}) with the Josephson energy equal to the collective component $E_J = E^c_J = |\beta|$. To compare the magnitude of the collective component of the Josephson energy $E_J^c$ with its individual counterpart given by
\begin{equation}
E_J^i = \bigl[\sum_{k=1}^K |\beta_k -\frac{1}{K}\beta |^2 \bigr]^{1/2} =\bigl[\sum_{k=1}^K |\beta_k|^2 -\frac{1}{K}|\beta |^2 \bigr]^{1/2}
\label{ind}
\end{equation}
one can consider a simple toy model with $\beta_k = Ae^{i\lambda k} , k= 0,1,...,K-1$. Then
\begin{equation}
E_J^c = |A|\frac{|1-e^{i\lambda K}|}{|1-e^{i\lambda}|}\leq \frac{2|A|}{|1-e^{i\lambda}|}\ ,\ E_J^i \simeq |A|\sqrt{K}
\label{ind1}
\end{equation}
what implies for a generic $\lambda$ that $E_J^i \sim E_J^c \sqrt{K}$.
\par
On the other hand for purely random amplitudes $\beta_k$,  $|\beta|^2 = \sum_k |\beta_k|^2$ and therefore $E_J^i \simeq E_J^c$. The real system should be placed between these two extremal cases of strong interference and random behavior what implies that the ratio $E_J^i / E_J^c $
increases as a certain positive power of $K$ leading to the domination of the individual coupling.
\par
For large JJ's with small $E_C$ this effect is suppressed by the fact that the typical level splitting for a fixed $j$ determined by the Coulomb repulsion is much smaller than the level splitting for different $j$'s given by the superconducting gap. For small junctions those energy scales are comparable and the individual tunneling prevails. This implies
that the matrix elements of $\hat{T}_{\mathrm{red}}$ between the vectors with the same $j$ are negligible in comparison
with the elements between vectors with $|j-j'|=1$.
Hence the only relevant matrix elements are the following
\begin{equation}
\langle r;1|\hat{T}_{\mathrm{red}}|0\rangle = \xi_r\ ,\  \langle s;0|\hat{T}_{\mathrm{red}}|1\rangle =\xi_{s}.
\label{tunhamred1}
\end{equation}
Using (\ref{tunhamred1}) one can derive the full effective Hamiltonian of the CPB including  (\ref{hamcpb1},\ref{tunhamred}) which is a direct sum of two similar terms
\begin{equation}
\hat{H}_{CPB}= \hat{H}^0_{CPB}\oplus\hat{H}^1_{CPB}
\label{sumham}
\end{equation}
acting on the subspaces $\mathcal{H}^0_{\mathrm{eff}}$,$\mathcal{H}^1_{\mathrm{eff}}$ spanned by $\{|0\rangle , |r;1\rangle , r =1,2,...,K-1\}$ and
$\{|1\rangle , |s;0\rangle , s =1,2,...,K-1\}$, respectively. To a large extend both "subsystems" can be treated separately and completely analogically. Depending on the parameters either $\mathcal{H}^0_{\mathrm{eff}}$ or $\mathcal{H}^1_{\mathrm{eff}}$ contains a ground state of $\hat{H}_{CPB}$. The ground state is a starting point of the controlled evolution in all experiments and as shown below both subspaces are invariant with respect to the Hamiltonian and controls. Hence, in the following,  only one subspace, say $\mathcal{H}^0_{\mathrm{eff}}$,  represents the accessible system.
\par
\emph{Charge qubit picture}
Defining a normalized vector
\begin{equation}
|\xi\rangle =(E_J)^{-1}\sum_{r=1}^{K-1}\xi_r|r;1\rangle,\ E_J = \Bigl(\sum_{r=1}^{K-1}|\xi_r|^2\Bigr)^{1/2}
\label{xi}
\end{equation}
one can introduce the vectors $|\pm\rangle$
\begin{equation}
|+\rangle = \cos\frac{\theta}{2}|\xi\rangle +\sin\frac{\theta}{2}|0\rangle\ ,\  |-\rangle = \cos\frac{\theta}{2}|0\rangle -\sin\frac{\theta}{2}|\xi\rangle\ ,
\label{eigenvec}
\end{equation}
where $\theta$ is defined by $\cos\theta=E(n_g)/\sqrt{E(n_g)^2 + E_J^2}$, $E(n_g) = 4E_C (1-2n_g)$.  The \emph{qubit observables} are given by
\begin{equation}
\hat{\sigma}^+=\frac{1}{2}\bigl(\hat{\sigma}^x + i\hat{\sigma}^y\bigr) = |+\rangle\langle -| \ ,\ \hat{\sigma}^z = |+\rangle\langle +| - |-\rangle\langle -|\ ,\ \hat{\sigma}^0 =|+\rangle\langle +| + |-\rangle\langle -|.
\label{qubit}
\end{equation}
Extracting from $\hat{H}^0_{CPB}$ the overall constant ${E(n_g)}/2 +4E_C n_g^2$
one obtains a new form of the Hamiltonian
\begin{equation}
\hat{H}^0_{\mathrm{CPB}} =  \frac{\omega}{2} \hat{\sigma}^z  + \frac{E(n_g)}{2}\hat{P}_0\ ,\ \omega =\sqrt{E(n_g)^2 + E_J^2},
\label{qubitham}
\end{equation}
with two eigenvectors $|\pm\rangle$ separated by the energy difference $\omega$ and the
third $(K-2)$-fold degenerated level corresponding to the projector $\hat{P}_0 == \sum_{r=1}^{K-1}|r;1\rangle\langle r;1| - |\xi\rangle\langle\xi| $ and the energy $E(n_g)/2$.
This third level lies always between  $|\pm\rangle$, as $-\omega/2 \leq E(n_g)/2 \leq \omega/2$.
\par
Notice that both parameters, $E_J$ and $g$ are related to the gap $\Delta$ and hence can be modified by the external magnetic field.
The time-dependent external control is performed by the coupling through the total electric charge operator $\hat{Q}$ and the total electric current
$\hat{J}=i[\hat{H}_{\mathrm{CPB}},\ \hat{Q} ]$ which possess a similar direct sum structure as the Hamiltonian. When restricted to $\mathcal{H}^0_{\mathrm{eff}}$ they read
\begin{equation}
\hat{Q}^0 = e\bigl[(\sin\theta) \hat{\sigma}^x - (\cos\theta )\hat{\sigma}^z  -\hat{\sigma}^0 -2\hat{P}_0\bigr] \ ,\ \hat{J}^0= 2e\,\omega (\sin\theta) \hat{\sigma}^y.
\label{echarge1}
\end{equation}
Obviously, if the system is completely isolated, then the \emph{qubit Hilbert space} spanned by $|\pm\rangle $ is invariant with respect to the Hamiltonian and the external control yielding the usual model of charge qubit.
The $(K-2)$-fold degenerated energy level corresponding to $\hat{P}_0$ and the lowest energy levels of the second subsystem become important when the coupling to an environment is discussed.
\par
\emph{Dissipation and decoherence processes}
 
There exist many hypothesis concerning the leading mechanism of dissipation and decoherence in CPB. Some of the mechanisms can be, in principle, reduced by a proper engineering and the others seem to be not effective enough to account for the experimental data. The presented model of individual tunneling which couples the ground state to excited Cooper pairs states allows for a new mechanism of dissipation due to excited Cooper pair - phonon coupling. As the typical CPB frequency $\omega$ lies
well below the Debye cut-off  the dissipation through the coupling to phonons is not suppressed like, for example, in the case of quantum dots. The corresponding acoustic wavelength is of the order of 0.1 $\mu m$, comparable to the size of the CPB and to the average distance between electrons forming a Cooper pair,  what implies rather individual coupling of different Cooper pair modes $|k\rangle$ than the collective one. It follows again that the value of $j$ is not conserved in the relaxation processes.  The detailed analysis of the model based on the Markovian master equation for the reduced density matrix $\hat{\rho}$ of the CPB will be presented in \cite{AM}. Here, the written below
Bloch equations for  the level occupation probabilities $p_{\pm} = \langle\pm|\hat{\rho}|\pm\rangle $ , $p_0 = \mathrm{Tr}(\hat{\rho}\hat{P}_0)$ and the qubit coherence $\alpha = \mathrm{Tr}(\hat{\rho}\hat{\sigma}^+)$ can be treated as a phenomenological description:
\begin{eqnarray}
\frac{dp_+}{dt} &=& -(\Gamma +\Gamma^{(+)}_W +\Gamma^{(+)}_E) p_+ ,
\nonumber \\
\frac{dp_0}{dt} &=& \Gamma^{(+)}_W  p_+ -\Gamma^{(-)}_{W} p_0 ,
\label{bloch} \\
\frac{dp_-}{dt} &=& \Gamma p_+  + \Gamma^{(-)}_{W} p_0 -\Gamma^{(-)}_E p_- ,\nonumber\\
\frac{d\alpha}{dt} &=& \bigl(i\omega -\frac{1}{2}[\Gamma +\Gamma^{(+)}_W +\Gamma^{(+)}_E+\Gamma^{(-)}_E]\bigr)\alpha \nonumber .
\end{eqnarray}
The decay rates describe the relaxation and  leakage processes depicted on the Fig.1. In the standard 2-level model only a single relaxation rate $1/T_1$ corresponding to  $\Gamma$ in (\ref{bloch}) is present, but the additional pure dephasing rate is added. Here pure dephasing is absent, because it cannot appear for a linear coupling to a bosonic bath (see the discussion in \cite{A}) and the additional decay of the coherence amplitude $\alpha$ is due to the probability leakage.

\begin{figure}[!ht]
\begin{center}
\includegraphics[width=0.8\textwidth]{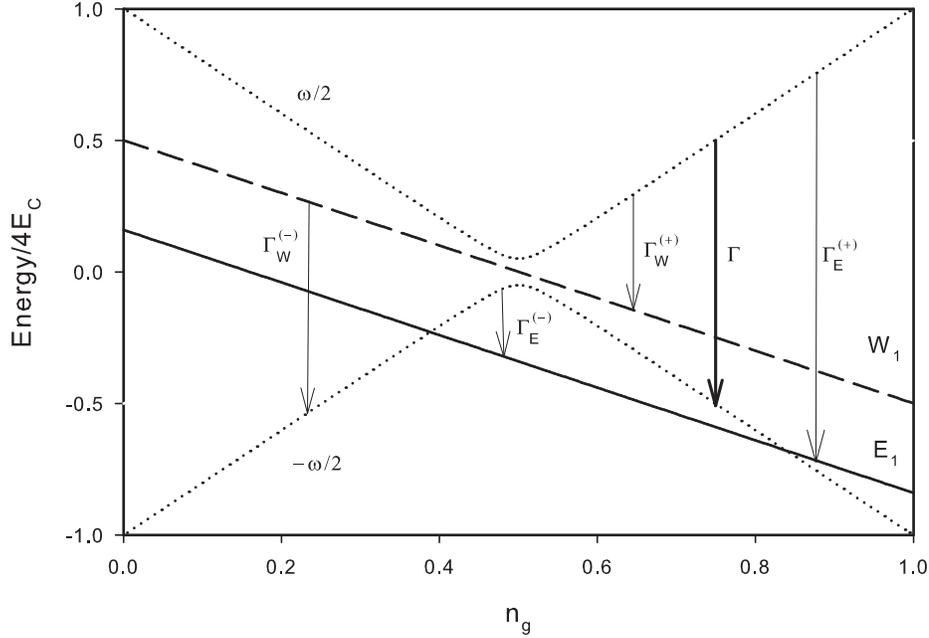}
\end{center}
\caption{Level structure and relaxation processes for CPB model,\\
 $^{......}$ - qubit levels , -- -- -- -- -- highly degenerated level.}
  \label{fig1}
\end{figure}

The Bloch equations (\ref{bloch}) and their extensions including a resonant electromagnetic perturbation will be used in \cite{AM} to analyse the experimental data, both for charge and phase qubits. Some observed effects like, for example, instability of the CPB ground state \cite{Leh} and the two time scales of energy relaxation for phase qubits \cite{Mar} will find an alternative explanations.
\par
\emph{Conclusions}
The idea that the collective tunneling of Cooper pairs is strongly suppressed by the interference, what for small Josephson junctions leads to the domination of an individual tunneling, yields a new model of a Cooper pair box. This model conceptually differs from the standard charge qubit picture but nevertheless reproduces  experimental data and even provides simple explanations for some issues. It suggests also the existence of a new effective relaxation mechanism due to the interaction of excited Cooper pairs with phonons. The fundamental difference between the standard and the new model is that the later cannot be treated as a model of \emph{macroscopic quantum system} with a single degree of freedom (e.g. large spin) but involves other degrees of freedom represented by highly degenerated states of excited Cooper pairs. This explains the absence of environmental effects leading to a semiclassical behavior for large quantum numbers.

\textbf{ Acknowledgments} The author thanks Frank Wilhelm and Wies\l aw Miklaszewski for discussions. This work is supported by the Polish Ministry of Science and Higher Education.

\end{document}